\documentstyle[12pt,axodraw]{article}
\newcommand{\JHEP}{J. High Energy Phys. }

\newcommand{\NP}{Nucl. Phys. }
\newcommand{\PR}{Phys. Rev. }

\newcommand{\PL}{Phys. Lett. }
\newcommand{\EPJ}{Eur. Phys. J. }
\newcommand{\IJMP}{Int. J. Mod. Phys. }

\newcommand{\pspace}{\int d^3\mu}
\newcommand{\bfq}{{\bf q}}
\newcommand{\bfk}{{\bf k}}
\newcommand{\bfp}{{\bf p}}

\begin{document}
\baselineskip=20pt

\pagenumbering{arabic}

\vspace{1.0cm}
\begin{flushright}
LU-ITP 2002/010
\end{flushright}

\begin{center}
{\Large\sf Time-ordered perturbation theory on noncommutative spacetime II: 
unitarity}\\[10pt]
\vspace{.5 cm}

{Yi Liao, Klaus Sibold}
\vspace{1.0ex}

{\small Institut f\"ur Theoretische Physik, Universit\"at Leipzig,
\\
Augustusplatz 10/11, D-04109 Leipzig, Germany\\}

\vspace{2.0ex}

{\bf Abstract}
\end{center}

We examine the unitarity issue in the recently proposed time-ordered 
perturbation theory on noncommutative (NC) spacetime. We show that unitarity 
is preserved as long as the interaction {\it Lagrangian} is explicitly 
Hermitian. We explain why it makes sense to distinguish the Hermiticity of the 
Lagrangian from that of the action in perturbative NC field theory and how this 
requirement fits in the framework. 

\begin{flushleft}
PACS: 02.40.Gh, 11.25.Db, 11.10.-z 

Keywords: noncommutative spacetime, time-ordered perturbation theory, 
unitarity 

\end{flushleft}

\newpage
\section{Introduction}

Quantum field theory on noncommutative (NC) spacetime has attracted a lot of 
activities since it was shown to appear as a limit of string theory in the 
presence of a constant NS-NS $B$ field background $\cite{string}$. New features 
in NC field theory have been found, such as the ultraviolet-infrared mixing 
$\cite{mixing}$, violation of unitarity $\cite{unitarity}\cite{follow}$ and 
causality $\cite{causality}$, which are very alien to ordinary field 
theory. These results are largely based on the understanding that field theory 
on NC spacetime may be formulated through the Moyal star product of functions 
on ordinary spacetime $\cite{doplicher}$ and that the only modification in 
perturbation theory is the appearance of momentum dependent NC phases at 
interaction vertices $\cite{filk}$. This naive approach in perturbation theory 
has been scrutinized recently in the context of the unitarity problem with the 
suggestion that the time-ordered product is not properly defined 
$\cite{bahns}$. It has also been shown explicitly that unitarity is preserved 
for the one-loop two-point function of $\varphi^3$ theory in the approach of 
Yang-Feldman equation $\cite{yfeldman}$. 

In a previous work $\cite{liao}$, we have reconsidered the issue of NC 
perturbation theory formulated in terms of the Moyal product. We assumed that 
perturbation theory can still be developped in the time-ordered expansion of a 
formally unitary time evolution operator specified by the interaction 
Lagrangian and that the usual concepts of time-ordering and commutation 
relations for free fields are still applicable. We found that the result is 
the old-fashioned, time-ordered perturbation theory (TOPT) 
$\cite{schweber}\cite{sterman}$ naturally extended to the NC case. In this 
framework, NC phases at interaction vertices are evaluated at on-shell momenta 
of positive or negative energy depending on the direction of time flow and are 
thus independent of the zero-th components of the generally off-shell momenta 
of participating particles. The analyticity properties of Green functions in 
the complex plane of the zero-th component are thus significantly modified. 
We explained how this in turn led to the result that this noncovariant 
formalism of TOPT cannot be recast into the seemingly covariant form of the 
naive approach when time does not commute with space. 
The whole picture of perturbation theory is thus altered; and 
this difference appears already at tree level in perturbation. It is then 
quite reasonable to ask whether some of the important statements made in the 
naive approach will be changed as well. In this work we address the issue of 
unitarity in the new framework and our conclusion on perturbative unitarity 
will indeed be different. 

In TOPT a process is described as a time sequence of transitions between 
physical intermediate states. The unitarity property in the TOPT formalism of 
ordinary field theory is usually transparent. Assuming that NC field theory is 
renormalizable at higher orders in perturbation theory, the unitarity proof 
in the TOPT framework for the usual field theory $\cite{sterman}$ almost goes 
through without change, up to one caveat which is specific to NC theory. 
Namely, the interaction {\it Lagrangian} must be explicitly Hermitian. While 
the Hermiticity of the Lagrangian guarantees that of the action, the opposite 
is not 
always automatic in NC field theory. In the naive approach one can appeal to 
the cyclicity of the spacetime integral of star products for the Hermiticity of 
the action even if one begins with a Lagrangian which is not explicitly 
Hermitian, e.g., 
${\cal L}_{\rm int}=-g\varphi^{\dagger}\star\varphi\star\sigma$.
However, as we stressed in Ref. $\cite{liao}$, it is important to notice 
that the time-ordering procedure does not commute with the star 
multiplication when time is involved in NC. Strictly insisting on this 
led to the conclusion that the naive, seemingly covariant approach of NC 
perturbation theory cannot be recovered from its TOPT formalism. 
Furthermore, the manipulation with star products in perturbation theory 
should not interfere with the time-ordering procedure. And as we shall show 
below, this has implications for the unitarity problem: to guarantee 
perturbative unitarity the interaction Lagrangian has to be explicitly 
Hermitian. This means that the above Lagrangian should be replaced by 
${\cal L}_{\rm int}=-g/2~
(\varphi^{\dagger}\star\varphi\star\sigma
+\sigma\star\varphi^{\dagger}\star\varphi)$.
While this is no harm to the action and quantities 
derived from it, it makes a big difference in the fate of unitarity in NC 
theory. 

In the next section we first make an {\it ab initio} calculation of the one 
loop contribution to the scalar self-energy. The purpose is to show 
explicitly that the prescriptions given in Ref. $\cite{liao}$ indeed apply 
as well to higher orders in perturbation. We confirm its unitarity as 
required for generally off-shell and amputated Green functions. This is 
followed by the study of a four-point function arising at one loop 
using the above prescriptions. Then, we show that it is necessary to make 
the interaction Lagrangian explicitly Hermitian to preserve the perturbative 
unitarity. This problem already appears at tree level as we shall illustrate 
by a four-point function. We explain how this requirement fits in the 
framework of TOPT. We summarize in the last section.

\section{Demonstration of perturbative unitarity}

We assume that perturbative field theory on NC spacetime formulated through 
the Moyal star product of field operators of ordinary spacetime functions
can still be developped in terms of vacuum expectation values of time-ordered 
products of field operators. The basic concepts such as time ordering and 
commutation relations for free fields are also assumed to be applicable. This 
is also the common starting point followed in the literature so far. Of course, 
there might be a drastic change with these assumptions in NC theory, but the 
philosophy is that we would like to avoid deviation from the well-established 
concepts as much as possible. But still, as we showed in the previous work, 
great differences amongst different approaches arise at a later stage when 
coping with time-ordered products. In this section, we provide one more 
difference concerning the fate of unitarity which is important for a theory to 
be consistent as a quantum theory. As we remarked in the Introduction, 
unitarity is almost obvious in the framework of TOPT on a formal level; 
however, an explicit demonstration 
of this in some examples is still instructive and interesting as we shall 
present below. Furthermore, it also leads to an observation concerning the 
Hermiticity of the interaction Lagrangian that has been ignored before in the 
context of NC field theory. 

\subsection{Self-energy of real scalar field}

Let us first study the scalar two-point function that is most frequently 
discussed in the literature on the unitarity issue. We consider the 
contribution arising from the following interaction, 
\begin{equation}
\begin{array}{rcl}
{\cal L}_{\rm int}&=&-g\left(\chi\star\varphi\star\pi+
\pi\star\varphi\star\chi\right), 
\end{array}
\label{eq_int1}
\end{equation}
where all fields are real scalars. We have deliberately used three different 
fields. On the one hand this makes our calculation more general than those 
considered so far, and on the other hand it avoids unnecessary complications 
arising from many possible contractions amongst identical fields which may be 
recovered later on by symmetrization. 

The one loop contribution to the $\varphi$ two-point function is 
\begin{equation}
\begin{array}{rcl}
G(x_1,x_2)&=&\displaystyle -\frac{g^2}{2!}\int d^4x_3\int d^4x_4
~A,\\
A&=&\displaystyle 
<0|T\left(\varphi_1\varphi_2
(\chi\star\varphi\star\pi+\pi\star\varphi\star\chi)_3 \right.\\
&&\displaystyle\times\left.
(\chi\star\varphi\star\pi+\pi\star\varphi\star\chi)_4 
\right)|0>, 
\end{array}
\label{eq_G}
\end{equation}
where from now on we use the indices of coordinates to specify the fields 
evaluated at corresponding points when no confusion arises. Our calculation 
is based on the following commutation relation between the positive- and 
negative-frequency parts of the field operator, e.g., $\varphi$,   
\begin{equation}
\begin{array}{rcl}
\varphi(x)&=&\varphi^+(x)+\varphi^-(x),\\
\left[\varphi^+(x),\varphi^-(y)\right]&=&\displaystyle D(x-y)\\ 
&=&\displaystyle 
\pspace_{\bf{p}}\exp[-ip_+\cdot (x-y)], 
\end{array}
\label{eq_comm}
\end{equation}
where $d^3\mu_{\bf{p}}=d^3{\bf p}[(2\pi)^3 2E_{\bf{p}}]^{-1}$ 
is the standard phase space measure with 
$E_{\bf{p}}=\sqrt{{\bf p}^2+m_{\varphi}^2}$, and 
$p_{\lambda}^{\mu}=(\lambda E_{\bf{p}},\bf{p})$ ($\lambda=\pm$) is the 
on-shell momentum with positive or negative energy. The calculation proceeds 
the same way as shown in Ref. $\cite{liao}$ although new complications arise 
due to the loop.  

We first compute the contractions of the $\chi$ and $\pi$ fields at 
interaction points $x_3$ and $x_4$. For $x_3^0>x_4^0$, only $\chi^-_4$ and 
$\pi_4^-$, which are on the right, and $\chi_3^+$ and $\pi_3^+$, which are on 
the left, can contribute. Shifting their positions using relations like 
eq. $(\ref{eq_comm})$ leads to the result, 
\begin{equation}
\begin{array}{rcl}
A&=&\displaystyle 
+<0|\cdots D_{34}(\chi)\star\varphi_3\cdots\varphi_4\star D_{34}(\pi)\cdots|0>\\ 
&&\displaystyle 
+\pspace_{\bfp}<0|\cdots e^{-ip_+^{\pi}\cdot x_3}\star\varphi_3\star 
D_{34}(\chi)\cdots\star\varphi_4\star e^{+ip_+^{\pi}\cdot x_4}|0>\\
&&\displaystyle 
+(\chi\leftrightarrow\pi), 
\end{array}
\end{equation}
where $D_{34}=D(x_3-x_4)$ and the argument or index $\chi$ ($\pi$) refers to 
the corresponding field and its mass being used. The dots represent possible 
positions for $\varphi_1$ and $\varphi_2$ fields appropriate to the time 
order. The first two terms originate from the diagonal and crossing contractions 
respectively, while the last arises because the interaction is symmetric in 
$\chi$ and $\pi$ fields. A little explanation on the star is necessary. 
Sometimes the same single $\star$ refers to both $x_3$ and $x_4$. This only 
means that the star multiplication is to be done separately with respect to 
$x_3$ and $x_4$. There never arises a case in which a star product is with 
respect to two different points since the only sourse of it is the interaction 
Lagrangian which is defined at a single point. This is a new feature at loop 
level that star products at different points get entangled. In principle 
this is not a problem and should not be confusing when we are careful enough, 
but it is a problem with notation. For example, the second term in the above 
equation may also be rewritten in a compact form as the first one; but we find 
that for this purpose we have to introduce more star symbols and specify which 
refers to which. This is even worse when more vertices are involved. For our 
aim of expressing the final result in momentum space, we find this is not 
worthwhile and it is much better to leave it as it stands. The result for the 
opposite case of $x_3^0<x_4^0$ is obtained by interchanging the indices $3$ and 
$4$. 

Next we contract the $\varphi$ fields. There are $4!$ time orders which we 
certainly do not have to consider one by one. They are classified into six 
groups, $T_{12}T_{34}$, $T_{34}T_{12}$, $T_{13}T_{24}$, $T_{24}T_{13}$, 
$T_{14}T_{23}$ and $T_{23}T_{14}$, where $T_{ij}T_{mn}$ stands for 
$(x_i^0{\rm ~and~}x_j^0)>(x_m^0{\rm ~and~}x_n^0)$. We only need to consider the 
first three groups while the last three can be obtained from the third by 
either $3\leftrightarrow 4$, or 
$1\leftrightarrow 2$, or both. The first two groups are relatively easy to 
compute. For example, for $x_1^0>x_2^0>x_3^0>x_4^0$ which belongs to one of 
the four possibilities in the first group, we have 
\begin{equation}
\begin{array}{rcl}
A&=&<0|\varphi_1\varphi_2\cdots\varphi_3\cdots\varphi_4\cdots|0>, 
\end{array}
\end{equation}
where the dots now represent the result from $\chi$ and $\pi$ contractions 
connected by stars which are computed above. There are actually four terms 
of course. Up to disconnected terms, $\varphi_{1,2}$ may be replaced by 
$\varphi_{1,2}^+$ and thus $\varphi_{3,4}$ by $\varphi_{3,4}^-$. 
Pushing further $\varphi_{1,2}^+$ to the right results in the following, 
\begin{equation}
\begin{array}{rcl}
A&=&\displaystyle 
\left[\right.
D_{34}(\chi)\star(D_{23}D_{14})\star D_{34}(\pi)\\ 
&&\displaystyle 
+\pspace_{\bfp}e^{-ip_+^{\pi}\cdot x_3}\star
(D_{23}\star D_{34}(\chi)\star D_{14})\star e^{+ip_+^{\pi}\cdot x_4}\\
&&\displaystyle 
+(1\leftrightarrow 2)\left.\right]
+(\chi\leftrightarrow\pi), 
\end{array}
\end{equation}
where the $D$ functions without an argument refer to the $\varphi$ field. The 
above is symmetric in $x_{1,2}$ and thus applies to the case of 
$(x_1^0{\rm ~and~}x_2^0)>x_3^0>x_4^0$. In this way we obtain the results for 
the first two groups of time orders, 
\begin{equation}
\begin{array}{rcl}
A&=&\displaystyle 
\tau_{34}\tau_{13}\tau_{23}\left\{\left[\right.\right.
D_{34}(\chi)\star(D_{23}D_{14})\star D_{34}(\pi)\\
&&\displaystyle 
+\pspace_{\bfp}e^{-ip_+^{\pi}\cdot x_3}\star 
(D_{23}\star D_{34}(\chi)\star D_{14})\star e^{+ip_+^{\pi}\cdot x_4}\\
&&\displaystyle 
+(1\leftrightarrow 2)\left.\right]
+(\chi\leftrightarrow\pi)\left.\right\}+(3\leftrightarrow 4), 
{\rm ~for~}T_{12}T_{34}, 
\end{array}
\end{equation}
\begin{equation}
\begin{array}{rcl}
A&=&\displaystyle 
\tau_{34}\tau_{41}\tau_{42}\left\{\left[\right.\right.
D_{34}(\chi)\star(D_{32}D_{41})\star D_{34}(\pi)\\
&&\displaystyle 
+\pspace_{\bfp}e^{-ip_+^{\pi}\cdot x_3}\star 
(D_{32}\star D_{34}(\chi)\star D_{41})\star e^{+ip_+^{\pi}\cdot x_4}\\
&&\displaystyle 
+(1\leftrightarrow 2)\left.\right]
+(\chi\leftrightarrow\pi)\left.\right\}+(3\leftrightarrow 4), 
{\rm ~for~}T_{34}T_{12}, 
\end{array}
\end{equation}
where $\tau_{jk}=\tau(x_j^0-x_k^0)$ is the step function.

The contractions for the case $T_{13}T_{24}$ is more complicated. Consider one 
of the four orders, $x_1^0>x_3^0>x_2^0>x_4^0$, for which we have the following 
structure, 
\begin{equation}
\begin{array}{rcl}
A&=&<0|\cdots\varphi_1\varphi_3\cdots\varphi_2\varphi_4\cdots|0>.
\end{array}
\end{equation}
Note that there is no problem for $\varphi_{1,2}$ to pass over the dots which 
contain the star products of c-number functions with respect to $x_{3,4}$. 
Then, $\varphi_1\varphi_3$ may be replaced by $(\varphi_1^+\varphi_3^++D_{13})$ 
and $\varphi_2\varphi_4$ by $(\varphi_2^-\varphi_4^-+D_{24})$. Up to 
disconnected terms, the remaining product of operators contributes a term 
$D_{32}\cdots D_{14}$ so that, 
\begin{equation}
\begin{array}{rcl}
A&=&\displaystyle 
D_{34}(\chi)\star(D_{13}D_{24}+D_{32}D_{14})\star D_{34}(\pi)\\
&&\displaystyle 
+\pspace_{\bfp}e^{-ip_+^{\pi}\cdot x_3}\star
(D_{13}\star D_{34}(\chi)\star D_{24}
+D_{32}\star D_{34}(\chi)\star D_{14})
\star e^{+ip_+^{\pi}\cdot x_4}\\
&&\displaystyle 
+(\chi\leftrightarrow\pi). 
\end{array}
\end{equation}
The complete sum for the case $T_{13}T_{24}$ can be put in a compact form, 
\begin{equation}
\begin{array}{rcl}
A&=&\displaystyle 
(\tau_{1324}+\tau_{1342}+\tau_{3124}+\tau_{3142})\\
&&\displaystyle\times\left\{\right.
D_{34}(\chi)\star(D_{32}D_{14})\star D_{34}(\pi)\\
&&\displaystyle 
+\pspace_{\bfp}e^{-ip_+^{\pi}\cdot x_3}\star(D_{32}\star D_{34}(\chi)
\star D_{14})\star e^{+ip_+^{\pi}\cdot x_4}
\left.\right\}\\
&&\displaystyle 
+\sum_{\lambda,\lambda^{\prime}}\tau_{13}^{\lambda}\tau_{(13)(24)}
\tau_{24}^{\lambda^{\prime}}\left\{\right.
D_{34}(\chi)\star(D_{13}^{\lambda}D_{24}^{\lambda^{\prime}})\star D_{34}(\pi)\\ 
&&\displaystyle 
+\pspace_{\bfp}e^{-ip_+^{\pi}\cdot x_3}\star(D_{13}^{\lambda}\star D_{34}(\chi) 
\star D_{24}^{\lambda^{\prime}})\star e^{+ip_+^{\pi}\cdot x_4}\left.\right\}\\
&&\displaystyle 
+(\chi\leftrightarrow\pi),
\end{array}
\label{eq_tau}
\end{equation}
where $\tau_{ijmn}=\tau_{ij}\tau_{jm}\tau_{mn}$, 
$\tau_{(ij)(mn)}=\tau(min(x_i^0,x_j^0)-max(x_m^0,x_n^0))$, and 
\begin{equation}
\begin{array}{l}
\tau_{jk}^{\lambda}=\left\{
                        \begin{array}{rl}
                        \tau_{jk},{\rm ~for~}\lambda=+\\
                        \tau_{kj},{\rm ~for~}\lambda=-
			\end{array}
		      \right.,
D_{jk}^{\lambda}=\left\{
                   \begin{array}{rl}
		   D_{jk},{\rm ~for~}\lambda=+\\
		   D_{kj},{\rm ~for~}\lambda=-
		   \end{array}		      
		   \right..  
\end{array}
\end{equation}

Now we show how to sum over 16 pairs of terms (not counting 
$\chi\leftrightarrow\pi$) so 
obtained in a desired form; namely, the connected contribution contains only 
functions $D_{13}^{\pm}$, $D_{14}^{\pm}$, $D_{23}^{\pm}$, $D_{24}^{\pm}$ and 
$D_{34}^{\pm}$ which should be accompanied by the corresponding step functions. 
We found four of them are already in the desired form. There are eight pairs, 
each of which is a combination of contributions from two time orders; for 
example, 
$\tau_{13}\tau_{23}\tau_{34}$ and $\tau_{13}\tau_{32}\tau_{24}$ 
unify precisely into the desired one $\tau_{13}\tau_{24}\tau_{34}$. 
Each of the remaining four pairs is again a combination of two contributions 
with one of them being of the same type as the first term in 
eq. $(\ref{eq_tau})$. 
They also unify comfortably into the desired time order; for example, the time 
order in the first term of eq. $(\ref{eq_tau})$ unifies with the one, 
$\tau_{32}\tau_{21}\tau_{14}$ from the other contribution, into 
$\tau_{14}\tau_{34}\tau_{32}$. 
Therefore, we have finally, 
\begin{equation}
\begin{array}{rcl}
A&=&\displaystyle 
+\sum_{\lambda_1,\lambda_2,\lambda}\left\{
\tau_{13}^{\lambda_1}\tau_{24}^{\lambda_2}\tau_{34}^{\lambda}\left[
D_{34}^{\lambda}(\chi)\star(D_{13}^{\lambda_1}D_{24}^{\lambda_2})\star 
D_{34}^{\lambda}(\pi)\right]
+(3\leftrightarrow 4)\right\}\\
&&\displaystyle 
+\sum_{\lambda_1,\lambda_2}\left\{\tau_{13}^{\lambda_1}
\tau_{24}^{\lambda_2}\right.
\pspace_{\bfp}\left[
e^{-ip_+^{\pi}\cdot x_3}\star 
(D_{13}^{\lambda_1}\star D_{34}(\chi)\star D_{24}^{\lambda_2})\star 
e^{+ip_+^{\pi}\cdot x_4}\right.\\
&&\displaystyle +\left.\left.
e^{-ip_+^{\pi}\cdot x_4}\star 
(D_{24}^{\lambda_2}\star D_{43}(\chi)\star D_{13}^{\lambda_1})\star 
e^{+ip_+^{\pi}\cdot x_3}\right]
+(3\leftrightarrow 4)\right\}\\
&&\displaystyle 
+(\chi\leftrightarrow\pi).
\end{array}
\end{equation}
Upon integrating over $x_{3,4}$, $(3\leftrightarrow 4)$ gives a factor of 
$2$ to cancel $1/2!$ in eq. $(\ref{eq_G})$ from the perturbation series, as 
expected. The trick to proceed further is the same as employed in Ref. 
$\cite{liao}$. Using 
\begin{equation}
\begin{array}{rcl}
\tau_{jk}^{\lambda}&=&\displaystyle 
\frac{i\lambda}{2\pi}\int_{-\infty}^{\infty}ds
\frac{\exp[-is(x_j^0-x_k^0)]}{s+i\epsilon\lambda},\\
D_{jk}^{\lambda}&=&\displaystyle 
\pspace_{\bfp}\exp[-ip_{\lambda}\cdot(x_j-x_k)],
\end{array}
\label{eq_rep}
\end{equation}
we can combine the $\tau$ function and its related three-momentum integral into 
a four-momentum integral. To make the result more symmetric in internal 
$\chi$ and $\pi$ lines, we may replace $\tau_{34}^{\lambda}$ by its square. We 
checked that the result is identical to the one using one factor of 
$\tau_{34}^{\lambda}$ as it must be as in ordinary field theory. We skip the 
further details and write down the result directly, 
\begin{equation}
\begin{array}{rcl}
G(x_1,x_2)&=&\displaystyle 
-g^2\int\frac{d^4p}{(2\pi)^4}\int\frac{d^4q}{(2\pi)^4}
\int\frac{d^4p_1}{(2\pi)^4}\int\frac{d^4p_2}{(2\pi)^4}
\sum_{\lambda_1,\lambda_2,\lambda}\\
&&\displaystyle \times 
iP_{\lambda}(p)iP_{\lambda}(q)iP_{\lambda_1}(p_1)iP_{\lambda_2}(p_2)
e^{-ip_1\cdot x_1}e^{-ip_2\cdot x_2}\\
&&\displaystyle \times 
(2\pi)^4\delta^4(p_1+p_2)(2\pi)^4\delta^4(p+q-p_1)({\rm NC~vertices}), 
\end{array}
\end{equation}
where $p,q,p_{1,2}$ refer to the $\pi,\chi,\varphi$ fields respectively so 
that their masses are implicit in on-shell quantities such as $E_{\bfp}$ and 
$p_{\lambda}$, and 
\begin{equation}
\begin{array}{rcl}
P_{\lambda}(k)&=&\displaystyle 
\frac{\lambda}{2E_{\bfk}[k^0-\lambda(E_{\bfk}-i\epsilon)]}\\
&=&\displaystyle \frac{\eta_{\lambda}(k)}{k^2-m^2+i\epsilon}, 
\end{array}
\end{equation}
with $\eta_{\lambda}(k)=1/2(1+\lambda k_0/E_{\bfk})$. 
The vertices have the factorized form, 
\begin{equation}
\begin{array}{rcl}
({\rm NC~vertices})&=&\displaystyle 
\left[e^{-i(q_{\lambda},-p_{1\lambda_1},p_{\lambda})}
     +e^{-i(p_{\lambda},-p_{1\lambda_1},q_{\lambda})}\right]\\
&\times &\displaystyle     
\left[e^{-i(q_{\lambda},+p_{2\lambda_2},p_{\lambda})}
     +e^{-i(p_{\lambda},+p_{2\lambda_2},q_{\lambda})}\right], 
\end{array}
\end{equation}
with 
$(k_1,k_2,\cdots,k_n)=\displaystyle 
\sum_{i<j}k_i\wedge k_j$ and $p\wedge q=1/2~\theta_{\mu\nu}p^{\mu}q^{\nu}$.  

Transforming into momentum space is now straightforward, 
\begin{equation}
\begin{array}{rcl}
\hat{G}(k_1,k_2)&=&\displaystyle 
\prod_{j=1}^2\left[\int d^4x_je^{-ik_j\cdot x_j}\right]G(x_1,x_2)\\
&=&\displaystyle 
-g^2(2\pi)^4\delta^4(k_1+k_2)\sum_{\lambda_1,\lambda_2}
iP_{\lambda_1}(k_1)iP_{\lambda_2}(k_2)\\
&&\displaystyle\times 
\sum_{\lambda}\int\frac{d^4p}{(2\pi)^4}\int\frac{d^4q}{(2\pi)^4}
(2\pi)^4\delta^4(p+q-k_1)iP_{\lambda}(p)iP_{\lambda}(q)\\
&&\times({\rm NC~vertices}), 
\end{array}
\end{equation}
where $k_{1,2}$ are the momenta flowing into the diagram. We have reversed 
the signs of variables $\lambda_j,\lambda$ and $p,q$ so that the only change 
in NC vertices is, 
$p_{1\lambda_1}\to k_{1\lambda_1}, 
 p_{2\lambda_2}\to k_{2\lambda_2}$. 
As we pointed out in Ref. $\cite{liao}$, it is important to notice that the 
zero-th components $p_0$ and $q_0$ are not involved in NC vertices which 
contain only on-shell momenta of positive or negative energy. This fact changes 
analyticity properties significantly. The $p_0,q_0$ integrals can thus be 
finished, one by the $\delta$ function, the other by a contour in its complex 
plane, with the result, 
\begin{equation}
\begin{array}{rl}
&i^{-1}\hat{G}(k_1,k_2)i^{-1}(k_1^2-m_{\varphi}^2)i^{-1}(k_2^2-m_{\varphi}^2)\\
=&\displaystyle 
-g^2(2\pi)^4\delta^4(k_1+k_2)\sum_{\lambda_1,\lambda_2}
\eta_{\lambda_1}(k_1)\eta_{\lambda_2}(k_2)\\
&\displaystyle\times 
\sum_{\lambda}\pspace_{\bfp}\pspace_{\bfq}(2\pi)^3\delta^3(\bfp+\bfq-\bfk_1)
[\lambda k_1^0-E_{\bfp}-E_{\bfq}+i\epsilon]^{-1}\\
&\times({\rm NC~vertices}), 
\end{array}
\end{equation}
which is exactly the amputated two-point function as can be obtained directly 
from the prescriptions given in Ref. $\cite{liao}$. 

We are now ready to examine the unitarity problem. We found that unitarity 
holds true in a detailed sense. Namely, as in ordinary field theory, it holds 
not only for the on-shell transition matrix but also for the off-shell 
amputated Green function. In the current noncovariant formalism, it even holds 
for separate configurations of external time direction parameters $\lambda_j$. 
This is not surprising since, if kinematically allowed, we can obtain the 
S-matrix elements for all possible channels of physical processes from the 
same Green function, which just correspond to different configurations of 
$\lambda_j$ and satisfy the unitarity relation. Let us check the following 
unitarity relation for the above example. Assuming 
$T(\{k_i,\lambda_i\}\to\{k_f,\lambda_f\})$ is the transition matrix or the 
amputated Green function for the process $i\to f$ with incoming momenta and 
time parameters $\{k_i,\lambda_i\}$ and outgoing ones $\{k_f,\lambda_f\}$, 
we have, 
\begin{equation}
\begin{array}{rl}
&\displaystyle 
-i\left[T(\{k_i,\lambda_i\}\to\{k_f,\lambda_f\})
     -T^*(\{k_f,\lambda_f\}\to\{k_i,\lambda_i\})\right]\\
=&\displaystyle 
\sum_n\prod_{j=1}^n\left[\pspace_{\bfp_j}\right]
T(\{k_i,\lambda_i\}\to n)T^*(\{k_f,\lambda_f\}\to n),
\end{array}
\label{eq_uni}
\end{equation}
where $n$ is a physical intermediate state with $n$ particles. 

Following the above convention, we change in our example 
$k_2\to -k_2$ and $\lambda_2\to -\lambda_2$ so that 
$k_{2\lambda_2}\to -k_{2\lambda_2}$, and 
the transition matrix becomes, 
\begin{equation}
\begin{array}{rl}
&T(\{k_1,\lambda_1\}\to\{k_2,\lambda_2\})\\
=&\displaystyle 
-g^2(2\pi)^4\delta^4(k_1-k_2)\sum_{\lambda}
\pspace_{\bfp}\pspace_{\bfq}(2\pi)^3\delta^3(\bfp+\bfq-\bfk_1)\\
&\displaystyle\times 
[\lambda k_1^0-E_{\bfp}-E_{\bfq}+i\epsilon]^{-1}V(1\to n)V(2\to n), 
\end{array}
\end{equation}
with $V(j\to n)=2\cos(q_{\lambda},-k_{j\lambda_j},p_{\lambda})$ being real so 
that only the physical threshold can develop an imaginary part, 
\begin{equation}
\begin{array}{rl}
&\displaystyle 
-i\left[T(\{k_1,\lambda_1\}\to\{k_2,\lambda_2\})
     -T^*(\{k_2,\lambda_2\}\to\{k_1,\lambda_1\})\right]\\
=&\displaystyle 
+g^2(2\pi)^4\delta^4(k_1-k_2)\sum_{\lambda}
\pspace_{\bfp}\pspace_{\bfq}(2\pi)^3\delta^3(\bfp+\bfq-\bfk_1)\\
&\displaystyle\times 
2\pi\delta(\lambda k_1^0-E_{\bfp}-E_{\bfq})V(1\to n)V(2\to n). 
\end{array}
\end{equation}
For a given sign of $k_1^0$, only one term in the sum over 
$\lambda$ actually contributes while the sum automatically includes both 
cases. For $k_1^0<0$, it is physically better to work with the inverse process. 
The above is precisely what we obtain for the right-hand side of eq. 
$(\ref{eq_uni})$ using the prescriptions for the two transitions; it is, for 
example, for $k_1^0>0$, 
\begin{equation}
\begin{array}{l}
\displaystyle\pspace_{\bfp}\pspace_{\bfq}\\
\displaystyle\times
\left[(2\pi)^3\delta^3(\bfk_1-\bfp-\bfq)(-2\pi)
\delta(k_1^0-E_{\bfp}-E_{\bfq})gV(1\to n)\right]\\
\displaystyle\times
\left[(2\pi)^3\delta^3(\bfp+\bfq-\bfk_2)(-2\pi)
\delta(E_{\bfp}+E_{\bfq}-k_2^0)gV(2\to n)\right]^*, 
\end{array}
\end{equation}
and unitarity is thus verified. For the complete and amputated Green function 
we merely have to multiply a real factor of $\eta_{\lambda_j}(k_j)$ for each 
external line and sum over $\lambda_j$. The on-shell transition matrix is 
projected by $k_j^0\to \lambda_j E_{\bfk_j}$. These manipulations do not 
lead to further problems. We note that the reality of NC vertices 
originating from the Hermiticity of ${\cal L}_{\rm int}$ in eq. 
$(\ref{eq_int1})$ plays an important role. Coping with a single real scalar 
field would not make this point so clear since ${\cal L}_{\rm int}$ would be 
automatically Hermitian. The latter case may be recovered by symmetrization 
which is already clear from our previous study. 

\subsection{Four-point function of real scalar field}

As a second example to demonstrate unitarity and to show applications of the 
prescriptions in Ref. $\cite{liao}$, we consider the $\varphi$ 
four-point function arising from the interaction, 
\begin{equation}
\begin{array}{rcl}
{\cal L}_{\rm int}&=&\displaystyle 
-g\varphi\star(\chi\star\pi+\pi\star\chi)\star\varphi,
\end{array}
\end{equation}
where all fields are real again. The lowest order contribution arises at one 
loop which has three Feynman diagrams. Here we shall consider only one of them, 
shown in Fig. $1$, while the other two may be obtained by permutation of 
indices. Their unitarity may be checked separately. Each Feynman diagram 
corresponds to two time-ordered diagrams which are represented collectively in 
Fig. $1$ by the parameter $\lambda=\pm$. For the configuration of incoming 
$k_{1,2},\lambda_{1,2}$ and outgoing $k_{3,4},\lambda_{3,4}$, we have 
\begin{equation}
\begin{array}{rcl}
T(12\to 34)&=&\displaystyle 
-2\pi\delta(k_1^0+k_2^0-k_3^0-k_4^0)\pspace_{\bfp}\pspace_{\bfq}
\sum_{\lambda}\\ 
&&\displaystyle\times 
\left[(2\pi)^3\delta^3(\bfk_1+\bfk_2-\bfp-\bfq)gV_{12}\right]\\
&&\displaystyle\times 
\left[(2\pi)^3\delta^3(\bfk_3+\bfk_4-\bfp-\bfq)gV_{34}\right]\\
&&\displaystyle\times 
\left[\lambda(k_1^0+k_2^0)-E_{\bfp}-E_{\bfq}+i\epsilon\right]^{-1}, 
\end{array}
\label{eq_uni2}
\end{equation}
where 
\begin{equation}
\begin{array}{rcl}
V_{12}&=&\left[
e^{-i(k_{1\lambda_1},-q_{\lambda},-p_{\lambda},k_{2\lambda_2})}
+(q_{\lambda}\leftrightarrow p_{\lambda})\right]
+(k_{1\lambda_1}\leftrightarrow k_{2\lambda_2}),\\
V_{34}&=&\left[
e^{-i(-k_{3\lambda_3},q_{\lambda},p_{\lambda},-k_{4\lambda_4})}
+(q_{\lambda}\leftrightarrow p_{\lambda})\right]
+(k_{3\lambda_3}\leftrightarrow k_{4\lambda_4}). 
\end{array}
\end{equation}
$(q_{\lambda}\leftrightarrow p_{\lambda})$ is due to the Hermitian 
arrangement of the $\chi$ and $\pi$ fields in ${\cal L}_{\rm int}$, while 
$(k_{1\lambda_1}\leftrightarrow k_{2\lambda_2})$ or 
$(k_{3\lambda_3}\leftrightarrow k_{4\lambda_4})$ is from symmetrization in the 
two $\varphi$ fields. We thus have, 
\begin{equation}
\begin{array}{rcl}
V_{12}&=&\displaystyle 
2^2\cos[k_{1\lambda_1}\wedge k_{2\lambda_2}+
(p_{\lambda}+q_{\lambda})\wedge(k_{1\lambda_1}-k_{2\lambda_2})]
\cos(p_{\lambda}\wedge q_{\lambda}), \\
V_{34}&=&\displaystyle 
2^2\cos[k_{3\lambda_3}\wedge k_{4\lambda_4}+
(p_{\lambda}+q_{\lambda})\wedge(k_{3\lambda_3}-k_{4\lambda_4})]
\cos(p_{\lambda}\wedge q_{\lambda}), 
\end{array}
\end{equation}
which are real again and do not contribute to the imaginary part of the 
transition matrix. For the $\varphi^4$ interaction of a single real scalar 
field, one merely has to symmetrize $V_{12}$ and $V_{34}$ further by including 
all permutations. 

\begin{center}
\begin{picture}(180,110)(0,0)
\SetOffset(0,10)
\ArrowArcn(80,50)(30,-180,0)
\ArrowArc(80,50)(30,-180,0)
\ArrowLine(20,70)(50,50)\ArrowLine(20,30)(50,50)
\ArrowLine(110,50)(140,70)\ArrowLine(110,50)(140,30)
\Text(80,90)[]{$\chi$}\Text(80,70)[]{$q,\lambda$}
\Text(80,10)[]{$\pi$}\Text(80,30)[]{$p,\lambda$}
\Text(15,70)[]{$\varphi$}\Text(35,75)[]{$k_1,\lambda_1$}
\Text(15,30)[]{$\varphi$}\Text(35,25)[]{$k_2,\lambda_2$}
\Text(145,70)[]{$\varphi$}\Text(125,75)[]{$k_3,\lambda_3$}
\Text(145,30)[]{$\varphi$}\Text(125,25)[]{$k_4,\lambda_4$}
\SetOffset(0,0)
\Text(90,0)[]{Fig. 1: Diagram corresponding to eq. $(\ref{eq_uni2})$.}
\end{picture}
\end{center}

The left-hand side of the unitarity relation is 
\begin{equation}
\begin{array}{rl}
&-i[T(12\to 34)-T^*(34\to 12)]\\
=&\displaystyle (-i)(-2\pi)(-i2\pi)
\delta(k_1^0+k_2^0-k_3^0-k_4^0)\pspace_{\bfp}\pspace_{\bfq}\sum_{\lambda}\\ 
&\displaystyle\times 
\left[(2\pi)^3\delta^3(\bfk_1+\bfk_2-\bfp-\bfq)gV_{12}\right]\\
&\displaystyle\times 
\left[(2\pi)^3\delta^3(\bfk_3+\bfk_4-\bfp-\bfq)gV_{34}\right]\\
&\displaystyle\times 
\delta\left(\lambda(k_1^0+k_2^0)-E_{\bfp}-E_{\bfq}\right), 
\end{array}
\end{equation}
which becomes, e.g., for $k_1^0+k_2^0>0$, 
\begin{equation}
\begin{array}{rl}
&\displaystyle 
\pspace_{\bfp}\pspace_{\bfq}
\left[-(2\pi)^4\delta^4(k_1+k_2-p_+-q_+)gV_{12}\right]\\
&\displaystyle\times 
\left[-(2\pi)^4\delta^4(p_++q_+-k_3-k_4)gV_{34}\right], 
\end{array}
\end{equation}
precisely the right-hand side of the unitarity relation. We have also checked 
that the above result is identical to that of an {\it ab initio} calculation. 

\subsection{Hermiticity of Lagrangian and unitarity of S-matrix}

From the above two examples it is already clear that the Hermiticity of 
the interaction Lagrangian is crucial to preserving unitarity as it is in 
ordinary theory. 
But they also seem to indicate that only real NC couplings are allowed for this 
purpose. If this were the case, NC field theory would be much less interesting 
as far as the standard model is concerned. We would like to clarify this point 
in this subsection; namely, there is no obstacle for a complex NC coupling to 
appear and the only requirement is the explicit Hermiticity of the interaction 
Lagrangian. We also point out the difference in the Hermiticity of the 
Lagrangian 
and action that is specific to perturbative NC field theory and that has not 
been noticed so far. In the meanwhile, we shall explain how this difference 
fits in the framework of TOPT. All of these points cannot be properly realized 
in $\varphi^3$ or $\varphi^4$ theory of a single real scalar that has been most 
frequently used in the literature in this context. 

Let us consider the following interaction Lagrangian, 
\begin{equation}
\begin{array}{rcl}
{\cal L}_{\rm int}^{\prime}&=&
-g_{\varphi}\varphi^{\dagger}\star\varphi\star\sigma
-g_{\chi}\chi^{\dagger}\star\chi\star\sigma, 
\end{array}
\label{eq_prob}
\end{equation}
where $\sigma$ ($\varphi,\chi$) is a real (complex) scalar and the coupling 
constants $g_{\varphi,\chi}$ are real. 
According to the understanding in the naive approach, the above is 
well defined in the sense that the action 
$\displaystyle S_{\rm int}^{\prime}=\int d^4x{\cal L}_{\rm int}^{\prime}$ 
is Hermitian by using the cyclicity property of integrals of star products, 
although ${\cal L}_{\rm int}^{\prime}$ is not in itself. The cyclicity 
argument in turn is based on integration by parts and ignoring surface terms. 
However, this integration by parts, when involving time derivatives, may 
clash with the time-ordering procedure in perturbation theory expanded in 
$S_{\rm int}^{\prime}$. Thus the above argument may break down in perturbation 
theory and cause problems. We have seen in Ref. $\cite{liao}$ a similar case 
of noncommutativity that the time-ordering procedure does not commute 
with star multiplication making the naive approach not recoverable from the 
TOPT framework when time does not commute with space. 

\begin{center}
\begin{picture}(180,120)(0,0)
\SetOffset(0,30)
\ArrowLine(40,40)(100,40)
\ArrowLine(10,10)(40,40)\ArrowLine(10,70)(40,40)
\ArrowLine(100,40)(130,10)\ArrowLine(100,40)(130,70)
\Text(70,50)[]{$\sigma$}\Text(70,30)[]{$p,\lambda$}
\Text(10,75)[]{$\varphi^{\dagger}$}\Text(35,65)[]{$k_2,\lambda_2$}
\Text(10,5)[]{$\varphi$}\Text(35,15)[]{$k_1,\lambda_1$}
\Text(130,75)[]{$\chi^{\dagger}$}\Text(105,65)[]{$k_4,\lambda_4$}
\Text(130,5)[]{$\chi$}\Text(105,15)[]{$k_3,\lambda_3$}
\SetOffset(0,0)
\Text(90,0)[]{Fig. 2: Diagram corresponding to eq. $(\ref{eq_uni3})$.}
\end{picture}
\end{center}

To see the unitarity problem resulting from eq. $(\ref{eq_prob})$, it is 
sufficient to consider the transition matrix for the following scattering at 
tree level, 
\begin{equation}
\begin{array}{rcl}
\varphi(k_1,\lambda_1)+\varphi^{\dagger}(k_2,\lambda_2)\to
\chi(k_3,\lambda_3)+\chi^{\dagger}(k_4,\lambda_4), 
\end{array}
\end{equation}
where $(k_{1,2},\lambda_{1,2})$ are incoming while $(k_{3,4},\lambda_{3,4})$ are 
outgoing. $k_j$'s are not necessarily on-shell, and $\lambda_j$'s are 
meaningful only when their connections to vertices are specified as we do in 
Fig. $2$. The $T$ matrix for a fixed configuration of $\lambda_j$'s is, 
\begin{equation}
\begin{array}{rcl}
T(12\to 34)&=&\displaystyle (-2\pi)\delta(k_1^0+k_2^0-k_3^0-k_4^0)
\pspace_{\bfp}\sum_{\lambda}\\
&&\displaystyle\times
\left[(2\pi)^3\delta^3(\bfk_1+\bfk_2-\bfp)g_{\varphi}V_{12}^{\prime}
\right]\\
&&\displaystyle\times
\left[(2\pi)^3\delta^3(\bfp-\bfk_3-\bfk_4)g_{\chi}V_{34}^{\prime}
\right]\\
&&\displaystyle\times
\left[\lambda(k_1^0+k_2^0)-E_{\bfp}+i\epsilon\right]^{-1}\\
&=&\displaystyle 
-(2\pi)^4\delta^4(k_1+k_2-k_3-k_4)\\
&&\displaystyle\times\sum_{\lambda} 
\frac{g_{\varphi}V_{12}^{\prime}g_{\chi}V_{34}^{\prime}}
{2E_{\bfp}[\lambda(k_1^0+k_2^0)-E_{\bfp}+i\epsilon]},
\end{array}
\label{eq_uni3}
\end{equation}
where $\bfp=\bfk_1+\bfk_2$, $E_{\bfp}=\sqrt{\bfp^2+m_{\sigma}^2}$ and similarly 
for other energies. The NC vertices are, 
\begin{equation}
\begin{array}{rcl}
V_{12}^{\prime}&=&\displaystyle 
\exp[-i(k_{2\lambda_2},k_{1\lambda_1},-p_{\lambda})],\\
V_{34}^{\prime}&=&\displaystyle 
\exp[-i(k_{3\lambda_3},k_{4\lambda_4},-p_{\lambda})], 
\end{array}
\end{equation}
where we have used $(-a,-b,c)=(a,b,-c)$ for $V_{34}^{\prime}$. For the inverse 
transition of incoming $(k_{3,4},\lambda_{3,4})$ and outgoing 
$(k_{1,2},\lambda_{1,2})$, we have, 
\begin{equation}
\begin{array}{rcl}
T(34\to 12)&=&\displaystyle (-2\pi)\delta(k_3^0+k_4^0-k_1^0-k_2^0)
\pspace_{\bfp}\sum_{\lambda}\\
&&\displaystyle\times
\left[(2\pi)^3\delta^3(\bfk_3+\bfk_4-\bfp)g_{\chi}\bar{V}_{34}^{\prime}
\right]\\
&&\displaystyle\times
\left[(2\pi)^3\delta^3(\bfp-\bfk_1-\bfk_2)g_{\varphi}\bar{V}_{12}^{\prime}
\right]\\
&&\displaystyle\times
\left[\lambda(k_3^0+k_4^0)-E_{\bfp}+i\epsilon\right]^{-1}\\
&=&\displaystyle 
-(2\pi)^4\delta^4(k_1+k_2-k_3-k_4)\\
&&\displaystyle\times\sum_{\lambda} 
\frac{g_{\varphi}\bar{V}_{12}^{\prime}g_{\chi}\bar{V}_{34}^{\prime}}
{2E_{\bfp}[\lambda(k_1^0+k_2^0)-E_{\bfp}+i\epsilon]},
\end{array}
\end{equation}
with 
\begin{equation}
\begin{array}{rcl}
\bar{V}_{12}^{\prime}&=&\displaystyle 
\exp[-i(k_{1\lambda_1},k_{2\lambda_2},-p_{\lambda})],\\
\bar{V}_{34}^{\prime}&=&\displaystyle 
\exp[-i(k_{4\lambda_4},k_{3\lambda_3},-p_{\lambda})]. 
\end{array}
\end{equation}
Noting that $V_{ij}^{\prime}\ne\bar{V}_{ij}^{\prime *}$, these factors will not 
factorize when 
forming the difference on the left-hand side of the unitarity relation, 
$-i\left[T(12\to 34)-T^*(34\to 12)\right]$. On the other hand, the right-hand 
side of the relation factorizes of course, 
\begin{equation}
\begin{array}{rl}
&\displaystyle 
\sum_{\lambda}\pspace_{\bfp}
\left[-(2\pi)\delta(k_1^0+k_2^0-\lambda E_{\bfp})
(2\pi)^3\delta^3(\bfk_1+\bfk_2-\bfp)g_{\varphi}V_{12}^{\prime}
\right]\\
&\displaystyle\times 
\left[-(2\pi)\delta(k_3^0+k_4^0-\lambda E_{\bfp})
(2\pi)^3\delta^3(\bfk_3+\bfk_4-\bfp)g_{\chi}\bar{V}_{34}^{\prime *}
\right]\\
=&\displaystyle 
(2\pi)^4\delta^4(k_1+k_2-k_3-k_4)\frac{1}{2E_{\bfp}}\sum_{\lambda}
2\pi\delta(k_1^0+k_2^0-\lambda E_{\bfp})
g_{\varphi}V_{12}^{\prime}g_{\chi}\bar{V}_{34}^{\prime *}. 
\end{array}
\end{equation}
Unitarity is thus violated at tree level.

Let us take a closer look at what goes wrong in the above. For example, using 
the spatial momentum conservation at the vertex, we have, 
\begin{equation}
\begin{array}{rl}
&2(k_{1\lambda_1},k_{2\lambda_2},-p_{\lambda})\\
=&\displaystyle 
\theta_{0i}\left[
(\lambda E_{\bfp}-\lambda_1 E_{\bfk_1}-\lambda_2 E_{\bfk_2})p^i+
(\lambda_2 E_{\bfk_2}k_1^i-\lambda_1 E_{\bfk_1}k_2^i)\right]
+\theta_{ij}k_1^ik_2^j. 
\end{array}
\end{equation}
Requiring $V_{jk}^{\prime}=\bar{V}_{jk}^{\prime *}$ 
($jk=12,34$), which guarantees unitarity, amounts to vanishing of the 
following, 
\begin{equation}
\begin{array}{rl}
&(k_{1\lambda_1},k_{2\lambda_2},-p_{\lambda})
  +(k_{2\lambda_2},k_{1\lambda_1},-p_{\lambda})\\
=&\displaystyle 
\theta_{0i}
(\lambda E_{\bfp}-\lambda_1 E_{\bfk_1}-\lambda_2 E_{\bfk_2})p^i, 
\end{array}
\label{eq_0}
\end{equation}
and similarly for $jk=34$. We make a few observations on the above result. 

First, space-space NC does not pose a problem with unitarity in the framework 
of TOPT. Namely, even if one starts with a Lagrangian such as eq. 
$(\ref{eq_prob})$ which can only be brought to be Hermitian by the cyclicity 
property, there will be no problem as long as $\theta_{0i}=0$. This is because 
the time-ordering procedure in perturbation theory does not interfere with the 
partial integration of spatial integrals employed in the cyclicity property. 
Furthermore, this freedom in partial integration corresponds exactly to the 
spatial momentum conservation at each separate vertex of TOPT. Conversely, we 
do not have such a freedom in temporal integration, which would spoil the 
time-ordering procedure, so that the temporal component of momentum does not 
conserve at each separate vertex in TOPT. But still we have a global 
conservation law for it which corresponds to the same amount of shift for all 
time parameters without disturbing their relative order. 

Second, for the particular example considered here, when all external particles 
are on-shell, we still have a chance to saturate unitarity even if 
$\theta_{0i}\ne 0$. For instance, when all $k_j^0=E_{\bfk_j}$ and 
$\lambda_j=+$, only $\lambda=+$ contributes to the unitarity relation so that 
unitarity holds true if 
$E_{\bfp}=E_{\bfk_1}+E_{\bfk_2}=E_{\bfk_3}+E_{\bfk_4}$.  
For the off-shell transition, which is a sum over all configurations of 
$\lambda_j$ and $\lambda$, eq. $(\ref{eq_0})$ cannot always vanish and thus 
there is no unitarity for the off-shell function. For transitions involving 
more than one internal line or loops so that we may have more freedom in 
spatial momenta of intermediate states, vanishing of similar combinations 
cannot be generally fulfilled. Thus we should not rely on this even for a 
cure to S-matrix unitarity. The same comment also applies to the 
kinematical configuration of $\theta_{0i}p^i=0$. 

The solution to this problem is already clear from the above discussion. 
Whenever time-space NC enters, we should make the interaction Lagrangian 
explicitly Hermitian before we do perturbation. In our example, instead of 
eq. $(\ref{eq_prob})$, we should start with the following one, 
\begin{equation}
\begin{array}{rcl}
{\cal L}_{\rm int}&=&\displaystyle 
-\frac{g_{\varphi}}{2}
(\varphi^{\dagger}\star\varphi\star\sigma
+\sigma\star\varphi^{\dagger}\star\varphi)\\
&&\displaystyle 
-\frac{g_{\chi}}{2}
(\chi^{\dagger}\star\chi\star\sigma
+\sigma\star\chi^{\dagger}\star\chi). 
\end{array}
\end{equation}
The only effect of this rearrangement is the substitution of the above primed 
vertices by the following ones, 
\begin{equation}
\begin{array}{rcl}
V_{12}&=&\displaystyle\frac{1}{2}\left(
 \exp[-i(k_{2\lambda_2},k_{1\lambda_1},-p_{\lambda})]
+\exp[-i(-p_{\lambda},k_{2\lambda_2},k_{1\lambda_1})]\right),\\
V_{34}&=&\displaystyle\frac{1}{2}\left(
 \exp[-i(k_{3\lambda_3},k_{4\lambda_4},-p_{\lambda})]
+\exp[-i(-p_{\lambda},k_{3\lambda_3},k_{4\lambda_4})]\right), 
\end{array}
\end{equation}
and then, 
\begin{equation}
\bar{V}_{12}=V_{12}^*,~\bar{V}_{34}=V_{34}^*, 
\end{equation}
which guarantees unitarity for any configurations since the difference 
on the left-hand side of the unitarity relation arises only from the physical 
threshold. 
More explicitly, we have, e.g., 
\begin{equation}
\begin{array}{rcl}
V_{12}&=&\displaystyle 
\exp(-ik_{2\lambda_2}\wedge k_{1\lambda_1})
\cos\left((k_{1\lambda_1}+k_{2\lambda_2})\wedge p_{\lambda}\right), 
\end{array}
\end{equation}
which is a complex coupling indeed. 

\begin{center}
\begin{picture}(350,110)(0,0)
\SetOffset(0,10)
\ArrowArcn(80,50)(30,-180,0)
\ArrowArc(80,50)(30,-180,0)
\ArrowLine(20,50)(50,50)
\ArrowLine(110,50)(140,50)
\Text(80,90)[]{$\chi$}\Text(80,70)[]{$q,\lambda$}
\Text(80,10)[]{$\pi^{\dagger}$}\Text(80,30)[]{$p,\lambda$}
\Text(15,50)[]{$\sigma$}\Text(35,60)[]{$k_1,\lambda_1$}
\Text(145,50)[]{$\rho$}\Text(125,60)[]{$k_2,\lambda_2$}

\SetOffset(175,10)
\ArrowArcn(80,50)(30,-180,0)
\ArrowArc(80,50)(30,-180,0)
\ArrowLine(20,50)(50,50)
\ArrowLine(110,50)(140,50)
\Text(80,90)[]{$\chi^{\dagger}$}\Text(80,70)[]{$q,\lambda$}
\Text(80,10)[]{$\pi$}\Text(80,30)[]{$p,\lambda$}
\Text(15,50)[]{$\sigma$}\Text(35,60)[]{$k_1,\lambda_1$}
\Text(145,50)[]{$\rho$}\Text(125,60)[]{$k_2,\lambda_2$}

\SetOffset(0,0)
\Text(175,0)[]{Fig. 3: Diagrams corresponding to eq. $(\ref{eq_uni4})$.}
\end{picture}
\end{center}

As a final example, we would like to illustrate the interplay between 
the Hermiticity of the Lagrangian and the contributions from complex conjugate 
intermediate states. We consider the one-loop induced $\sigma\to\rho$ 
transition through the following interactions, 
\begin{equation}
\begin{array}{rcl}
{\cal L}_{\rm int}&=&\displaystyle 
-(g_{\sigma}\chi^{\dagger}\star\pi\star\sigma
+g_{\sigma}^*\sigma\star\pi^{\dagger}\star\chi)
-(g_{\rho}\chi\star\pi^{\dagger}\star\rho
+g_{\rho}^*\rho\star\pi\star\chi^{\dagger}),                    
\end{array}
\end{equation}
where $\chi,\pi$ are complex scalars and $\rho,\sigma$ are real ones with 
generally complex couplings $g_{\sigma},g_{\rho}$. There are two Feynman 
diagrams with conjugate virtual particle pairs $\chi\pi^{\dagger}$ and 
$\chi^{\dagger}\pi$ respectively, and each of them has two time-ordered 
diagrams depicted collectively in Fig. $3$. We write down their 
contributions directly, 
\begin{equation}
\begin{array}{rl}
&T_{\chi\pi^{\dagger}}
(\sigma(k_1,\lambda_1)\to\rho(k_2,\lambda_2))\\
=&\displaystyle 
-(2\pi)^4\delta^4(k_1-k_2)g_{\sigma}g_{\rho}\pspace_{\bfp}\pspace_{\bfq}
(2\pi)^3\delta^3(\bfk_1-\bfp-\bfq)\\
&\displaystyle \times
\sum_{\lambda}\frac{V_{\sigma}V_{\rho}}
{\lambda k_1^0-E_{\bfp}-E_{\bfq}+i\epsilon},\\ 
&T_{\chi^{\dagger}\pi}
(\sigma(k_1,\lambda_1)\to\rho(k_2,\lambda_2))\\
=&\displaystyle 
-(2\pi)^4\delta^4(k_1-k_2)g_{\sigma}^*g_{\rho}^*
\pspace_{\bfp}\pspace_{\bfq}
(2\pi)^3\delta^3(\bfk_1-\bfp-\bfq)\\
&\displaystyle \times
\sum_{\lambda}\frac{V_{\sigma}^*V_{\rho}^*}
{\lambda k_1^0-E_{\bfp}-E_{\bfq}+i\epsilon}, 
\end{array}
\label{eq_uni4}
\end{equation}
with 
$V_{\sigma}=\exp[-i(q_{\lambda},p_{\lambda},-k_{1\lambda_1})]$, 
$V_{\rho}  =\exp[-i(q_{\lambda},p_{\lambda},-k_{2\lambda_2})]$. 
Similarly, for the inverse transition $\rho\to\sigma$, we have, 
\begin{equation}
\begin{array}{rl}
&T_{\chi\pi^{\dagger}}
(\rho(k_2,\lambda_2)\to\sigma(k_1,\lambda_1))\\
=&\displaystyle 
-(2\pi)^4\delta^4(k_1-k_2)g_{\sigma}^*g_{\rho}^*
\pspace_{\bfp}\pspace_{\bfq}
(2\pi)^3\delta^3(\bfk_1-\bfp-\bfq)\\
&\displaystyle \times
\sum_{\lambda}\frac{V_{\sigma}^*V_{\rho}^*}
{\lambda k_1^0-E_{\bfp}-E_{\bfq}+i\epsilon},\\ 
&T_{\chi^{\dagger}\pi}
(\rho(k_2,\lambda_2)\to\sigma(k_1,\lambda_1))\\
=&\displaystyle 
-(2\pi)^4\delta^4(k_1-k_2)g_{\sigma}g_{\rho}
\pspace_{\bfp}\pspace_{\bfq}
(2\pi)^3\delta^3(\bfk_1-\bfp-\bfq)\\
&\displaystyle \times
\sum_{\lambda}\frac{V_{\sigma}V_{\rho}}
{\lambda k_1^0-E_{\bfp}-E_{\bfq}+i\epsilon}. 
\end{array}
\end{equation}
Thus, with $g_{\sigma}$ and $g_{\rho}$ coupling terms alone (or their 
conjugate terms alone) it is impossible to fulfil unitarity because 
even the action is not Hermitian. If the action is Hermitian, it is 
always possible to make the Lagrangian Hermitian too. Once this is done, 
unitarity holds individually for the two conjugate intermediate states of  
$\chi\pi^{\dagger}$ and $\chi^{\dagger}\pi$.
This is precisely the same phenomenon occurring in ordinary field theory 
as may be easily checked for the above example. 

\section{Conclusion}

In a previous paper we proposed a framework to do perturbation theory for NC 
field theory which is essentially the time-ordered perturbation theory 
extended to the NC case. In contrast to ordinary field theory, this framework 
is not equivalent to the naive, seemingly covariant one pursued in the 
literature due to the significant change of analyticity properties introduced 
by NC phases. In the present paper we examined the impact of this change on the 
unitarity problem ocurring in the naive approach when time does not commute 
with space, and arrived at a different result on the fate of unitarity. Our 
main conclusion is that there is no problem with unitarity in TOPT as long as 
the interaction {\it Lagrangian} is explicitly Hermitian. We showed this 
explicitly in examples and then extended that result. 

The key observation in distinguishing the Hermiticity of the Lagrangian and 
that of the action in NC field theory is that the manipulation with the 
cyclicity property in spacetime integrals of star products may clash with the  
time-ordering procedure in perturbation theory. In ordinary theory there is no 
similar problem arising from integration by parts in the action because, even 
if one takes it seriously at the beginning, one can always remove it by going 
back to the covariant formalism by analytic continuation. However, in NC theory 
with time-space NC, 
as we argued previously, this continuation is not possible at least in the 
naive sense. It thus makes difference whether the Lagrangian is explicitly 
Hermitian or not. But we would like to stress again that requiring Hermiticity 
of the Lagrangian does not forbid complex NC couplings to appear. 

The main drawback of TOPT, as it is in ordinary theory, is its rapidly 
increased technical complication when going to higher orders in perturbation. 
It would be highly desirable if it could be recast in a more or less 
covariant form. 

{\bf Acknowledgements}

Y.L. would like to thank M. Chaichian for a visit at the Helsinki 
Institute of Physics and its members for hospitality. He enjoyed many 
encouraging discussions with M. Chaichian, P. Presnajder and A. Tureanu.
K.S. is grateful to D. Bahns and K. Fredenhagen for clarifying 
discussions on their work.


\begin{thebibliography}{30}

\bibitem{string}
A. Connes, M. R. Douglas and A. Schwarz, 
{\it Noncommutative geometry and matrix theory: compactification on tori}, 
\JHEP 02 (1998) 003 [hep-th/9711162]; 
M. R. Douglas and C. Hull, 
{\it D-branes and the noncommutative torus}, 
{\it ibid}. 02 (1998) 008 [hep-th/9711165];
C.-S. Chu and P.-M. Ho, 
{\it Noncommutative open string and D-brane}, 
\NP B550 (1999) 151 [hep-th/9812219]; 
{\it Constrained quantization of open string in background $B$ field and 
noncommutative D-brane}, 
{\it ibid}. B568 (2000) 447 [hep-th/9906192]; 
V. Schomerus, 
{\it D-branes and deformation quantization}, \JHEP 06 (1999) 030 
[hep-th/9903205].
N. Seiberg and E. Witten, 
{\it String theory and noncommutative geometry}, 
{\it ibid}. 09 (1999) 032 [hep-th/9908142].

\bibitem{mixing}
S. Minwalla, M. V. Raamsdonk and N. Seiberg, 
{\it Noncommutative perturbative dynamics}, 
\JHEP 02 (2000) 020 [hep-th/9912072];
I. Ya. Aref'eva, D. M. Belov and A. S. Koshelev, 
{\it Two-loop diagrams in noncommutative $\phi_4^4$ theory}, 
\PL B476 (2000) 431 [hep-th/9912075]; 
M. V. Raamsdonk and N. Seiberg, 
{\it Comments on noncommutative perturbative dynamics}, 
\JHEP 03 (2000) 035 [hep-th/0002186];
A. Matusis, L. Susskind and N. Toumbas, 
{\it The UV/IR connection in the noncommutative gauge theories},
\JHEP 12 (2000) 002 [hep-th/0002075].

\bibitem{unitarity}
J. Gomis and T. Mehen, 
{\it Space-time noncommutative field theories and unitarity}, 
\NP B591 (2000) 265 [hep-th/0005129]. 

\bibitem{follow}For subsequent discussions on unitarity, see for example: 
J. Gomis, K. Kamimura and J. Llosa, 
{\it Hamiltonian formalism for space-time noncommutative theories}, 
\PR D63 (2001) 045003 [hep-th/0006235]; 
O. Aharony, J. Gomis and T. Mehen, 
{\it On theories with light-like noncommutativity}, 
\JHEP 09 (2000) 023 [hep-th/0006236]; 
M. Chaichian, A. Demichev, P. Presnajder and A. Tureanu, 
{\it Space-time noncommutativity, discreteness of time and unitarity}, 
\EPJ C20 (2001) 767 [0007156]; 
R.-G. Cai and N. Ohta, 
{\it Lorentz transformation and light-like noncommutative SYM}, 
\JHEP 10 (2000) 036 [hep-th/0008119]; 
L. Alvarez-Gaume, J. L. F. Barbon, R. Zwicky, 
{\it Remarks on time-space noncommutative field theories}, 
\JHEP 05 (2001) 057 [hep-th/0103069]; 
T. Mateos, A. Moreno, 
{\it A note on unitarity of nonrelativistic noncommutative theories}, 
\PR D64 (2001) 047703 [hep-th/0104167]; 
A. Bassetto, L. Griguolo, G. Nardelli and F. Vian, 
{\it On the unitarity of quantum gauge theories on noncommutative spaces}, 
\JHEP 07 (2001) 008 [hep-th/0105257]; 
C.-S. Chu, J. Lukierski and W. J. Zakrzewski, 
{\it Hermitian analyticity, IR/UV mixing and unitarity of noncommutative 
field theories}, hep-th/0201144. 

\bibitem{causality}N. Seiberg, L. Susskind and N. Toumbas, 
{\it Space-time noncommutativity and causality}, 
\JHEP 06 (2000) 044 [hep-th/0005015]; 
L. Alvarez-Gaume and J.L.F. Barbon, 
{\it Nonlinear vacuum phenomena in noncommutative QED}, 
\IJMP A16 (2001) 1123 [hep-th/0006209].

\bibitem{doplicher}S. Doplicher, K. Fredenhagen and J. E. Roberts, 
{\it The quantum structure of spacetime at the Planck scale and quantum 
fields}, Commun. Math. Phys. 172 (1995) 187.

\bibitem{filk}T. Filk, {\it Divergences in a field theory on quantum space}, 
\PL B376 (1996) 53.

\bibitem{bahns}D. Bahns, S. Doplicher, K. Fredenhagen and G. Piacitelli, 
{\it On the unitarity problem in space-time noncommutative theories}, 
\PL B533 (2002) 178 [hep-th/0201222]; see also: 
C. Rim and J. H. Yee, 
{\it Unitarity in space-time noncommutative field theories}, hep-th/0205193.

\bibitem{yfeldman}C. N. Yang and D. Feldman, 
{\it The S matrix in the Heisenberg representation}, 
\PR 79 (1950) 972.

\bibitem{liao}Y. Liao and K. Sibold, 
{\it Time-ordered perturbation theory on noncommutative spacetime: 
basic rules}, hep-th/0205269.

\bibitem{schweber}S. S. Schweber, 
{\it An introduction to relativistic quantum field theory}, 
Harper $\&$ Row, 1961.

\bibitem{sterman}G. Sterman, 
{\it An introduction to quantum field theory}, 
Cambridge University Press, 1993.

\end{thebibliography}
\end{document}